# SYNCHROSCAN STREAK CAMERA IMAGING AT A 15-MEV PHOTOINJECTOR WITH EMITTANCE EXCHANGE*

A.H. Lumpkin, J. Ruan, and R. Thurman-Keup

Fermilab, Batavia, IL U.S.A. 60510

*Abstract*

At the Fermilab A0 photoinjector facility, bunch-length measurements of the laser micropulse and the e-beam micropulse have been done in the past with a fast single-sweep module of the Hamamatsu C5680 streak camera with an intrinsic shot-to-shot trigger jitter of 10 to 20 ps. We have upgraded the camera system with the synchroscan module tuned to 81.25 MHz to provide synchronous summing capability with less than 1.5-ps FWHM trigger jitter and a phase-locked delay box to provide phase stability of ~1 ps over 10s of minutes. These steps allowed us to measure both the UV laser pulse train at 263 nm and the e-beam via optical transition radiation (OTR). Due to the low electron beam energies and OTR signals, we typically summed over 50 micropulses with 0.25-1 nC per micropulse. The phase-locked delay box allowed us to assess chromatic temporal effects and instigated another upgrade to an all-mirror input optics barrel. In addition, we added a slow sweep horizontal deflection plug-in unit to provide dual-sweep capability for the streak camera. We report on a series of measurements made during the commissioning of these upgrades including bunch-length and phase effects using the emittance exchange beamline and simultaneous imaging of a UV drive laser component, OTR, and the 800-nm diagnostics laser.

## 1. INTRODUCTION

In analogy to upgrades performed two decades ago in support of rf linac driven free-electron lasers[1], the opportunity for a new series of streak camera experiments at the Fermilab A0 photoinjector (A0PI) on the 14-16 MeV electron beams and the UV component of the drive laser has been identified and now realized. These bunch-length and phase-stability measurements were driven by the need to diagnose accurately the beam's longitudinal emittance for the ongoing emittance exchange (EEX) studies at the facility [2-4]. The enabling upgrade was adding the synchroscan plug-in option to the existing C5680 Hamamatsu streak camera mainframe. By locking this module to the 81.25 MHz subharmonic of the rf system, the synchronous summing of micropulses could be done with trigger jitter of <1.5 ps (FWHM) for both the UV drive laser component at 263 nm and the e-beam via optical transition radiation (OTR) measurements [2,4]. This jitter is significantly lower than the 10-20 ps trigger jitter found in a fast single-sweep unit which precluded direct summing of the sub-10 ps pulses in the past. The synchronous summing of the low OTR signal from the 15-MeV electron beam micropulses allowed the needed bandpass filters to be utilized to reduce the chromatic temporal dispersion effects inherent to the broadband OTR source and the transmissive optics components. In addition, the C6768 delay module with phase feedback was also implemented, and this unit stabilized the streak camera sweep relative to the master oscillator so that camera phase drift was much reduced to the ps level over 10s of minutes. This latter feature allowed a series of experiments to be done on the bandwidth effects and transit-time effects in the respective transport lines which require longer term phase stability. After characterizing the UV laser bunch length, a series of e-beam experiments on the A0 beamlines was performed. In the course of our experiments, we did a series of tests on the chromatic temporal dispersion effects for this particular input optics barrel with UV transmitting optics and our optical transport lines. We show our effects were less than that reported at Stanford Synchrotron Radiation Lab with optical synchrotron radiation (OSR) [5], but ours still had to be characterized carefully to allow accurate bunch-length measurements using OTR. We have now installed an all-mirror input optics barrel to mitigate the effects. In addition, we obtained and implemented the slow horizontal sweep plug-in unit that allowed dual-sweep measurements of the micropulse phase and bunch lengths during the beam macropulse. We also describe the use of an ultrafast Ti:Sa laser to characterize the streak-tube temporal response, and we report initial steps in synchronizing this laser pulse with the drive laser and electron beam via OTR.

___

*Work supported by U.S. Department of Energy, Office of Science, Office of High Energy Physics, under Contract No. DE-AC02-07CH1135.

## 2. EXPERIMENTAL CONSIDERATIONS

The tests were performed at the Fermilab A0 photoinjector facility which includes an L-band photocathode (PC) rf gun and a 9-cell SC rf accelerating structure which combine to generate up to 16-MeV electron beams [2]. The drive laser operates at 81.25 MHz although the micropulse structure is usually counted down to 1 MHz. Previous bunch length measurements of the drive laser and e-beam [6, 7] were done with the fast single-sweep module of the Hamamatsu C5680 streak camera with an inherent shot-to-shot trigger jitter of 10 to 20 ps. Such jitter precluded synchronous summing of the short pulses. We have upgraded the camera by acquiring the M5676 synchroscan module tuned to 81.25 MHz with a trigger jitter of less than 1.5 ps (FWHM) and the C6878 phase-locked delay unit which stabilizes the camera phase over 10s of minutes. Due to the low, electron-beam energies and OTR signals, we typically synchronously summed over 50 micropulses with 1 nC per micropulse. The initial tests were performed in the straight-ahead line where energizing a dipole sends the beam into a final beam dump. The setup includes the upstream corrector magnets, quadrupoles, rf BPM, the YAG:Ce/OTR imaging stations, and the beam dump as schematically shown in Fig. 1. The initial sampling station was chosen at X9, and an optical transport system using flat mirrors and a parabolic mirror brought the light to the streak camera. A short focal length quartz lens was used to focus the beam image more tightly onto the streak camera entrance slit. The quartz-based UV-Vis input optics barrel transferred the slit image to the Hamamatsu C5680 streak camera's photocathode. One of the upgrades was to change this optics to all mirror input optics.

Alternatively, the four dipoles of the emittance exchange line could be powered and experiments done at an OTR station, X24. After the fourth dipole a second optical transport line brings the OTR to the streak camera. In the EEX line the bunch compression effects were observed, and the shorter bunches were used to help delineate the chromatic temporal dispersion effects for various band pass (BP), long pass (LP), and short pass (SP) filters. The OTR converter is an Al-coated optics mirror with a 1.5-mm-thick zerodur substrate or a 0.25-mm- thick Si wafer substrate, and it is mounted with its surface normal at 45 degrees to the beam direction on a stepper assembly. The assembly provides vertical positioning with an option for a YAG:Ce scintillator crystal position. A two-position actuator and a 4-position translation stage were used in the optical path in front of the camera to select band pass filters. The OTR streak readout camera images were recorded with a PCI-based video digitizer for both online and offline image analyses. The charge was monitored by an upstream current monitor. We describe a phased approach over time for our streak camera upgrades and characterizations in the following sections.

## 2.1 RESOLUTION AND BANDWIDTH EFFECTS

One of the first steps in verifying streak camera operations with OTR is to determine the static spread function contribution to temporal resolution. This is the vertical beam spot size of the entrance slit mapped through the imaging system when in streak camera "focus" mode. The major contribution is the slit height itself for values larger than 30 µm or so. In our early experiments with 10-50 nC of charge integrated in the micropulse sum, we used a slit height of 80 µm, which resulted in a limiting vertical spot size of 9 pixels. The limiting resolution is then found by multiplying this by the sweep-speed calibration factor. We did a careful determination of the two fastest ranges, range 2 and range 1 by using a laser pulse-stacker configuration. By splitting the laser beam energy, we could separately delay one pulse relative to the other by a set of movable mirrors. We then tracked the observed pulse separations in the streak camera images. Plots of the observed time separations are shown in Fig. 2 for range 2 (top) and for range 1 (bottom). The reciprocals of the fitted slopes gave us 1.55 ps/pixel and 0.32 ps/pixel, respectively. This means our initial resolution terms were 14.0 ps and 2.9 ps (FWHM), respectively. In the second series of experiments we reduced the slit height to 40 µm with a corresponding vertical spot size of 4.7 pixels (FWHM). This then gives us resolution terms of 7.3 ps and 1.50 ps (FWHM), respectively, for range 2 and range 1. This was needed for the bunch compression tests particularly.

One of the practical issues we addressed was the chromatic temporal dispersion that occurred for the broadband OTR light as it was transported through the transmissive components of the optical transport line. Since the input optics barrel of the streak camera initially was actually UV transmitting, it consisted of quartz optical components. This material has less variation of index of refraction with wavelength than flint glass or other materials used in the other standard Hamamatsu input optics, but still results in a measurable effect that limits effective temporal resolution with broadband light. Our effect was shown to be smaller than the Stanford Synchrotron Radiation Lab (SSRL) setup of 0.2 ps/nm reported previously [5]. The basic concept is expressed by the simple relationship for the transit-time change, $\Delta t = L (v_{g2}-v_{g1})/(v_{g1} \times v_{g2})$, due to the difference in group velocities $v_{g1}$ and $v_{g2}$ for two wavelengths through a characteristic material thickness, L [8].

This effect is represented in Fig. 3 where a 3-ps (FWHM) initial pulse is shown as arriving at different times for different wavelengths with a 4-ps shift across the bandwidth of the measurement. The resulting superposition of these Gaussian profiles can be fit to a single Gaussian of 4.21 ps (FWHM). In the actual MATLAB model, a series of over 1000 Gaussians was used. In our case the temporal shift was 8 to 9 ps within the 550-nm SP filter bandwidth and caused an effective limiting resolution term of about 4.4 pixels (FWHM) for range 2 in quadrature with the static spread function of 4.7 pixels. When we use a 550-nm LP filter this term is reduced further to 1.76 pixels in Range 2 and 8.8 pixels in Range 1.

We then can calculate the actual pulse length by subtracting from the total observed pulse width in pixels the contributing terms of static spread function, bandwidth, and trigger jitter. Since the jitter term appears to be small compared to our bunch lengths, we have absorbed it into the actual bunch length term for the time being. If we use a 550-nm longpass filter then for range 2 and range 1 we would have respectively:

$$\delta t(FWHM) = \sqrt{Pixel^2 - 4.7^2 - 1.76^2} \times 1.55\, ps/pixel \quad \text{Eqn. 1}$$

$$\delta t(FWHM) = \sqrt{Pixel^2 - 4.7^2 - 8.8^2} \times 0.32\, ps/pixel \quad \text{Eqn. 2}$$

The simple statistical uncertainty in the mean value, $\sigma_{stat}$ can be found by dividing the standard deviation, $s$, by the square root of the number of images processed, $N$, so $\sigma_{stat} = s/N^{1/2}$ [9]. However, since we have a nonlinear relation in Eq. 2, the total error, $s_t$, should also include the errors from the resolution term and the bandwidth term added in quadrature. For most bunch lengths encountered we can neglect the camera jitter term which is < 0.2 ps during a macropulse. The general equation for the error $s_t$ for the nonlinear relationship for $y(x_1, x_2, x_3, x_i)$ is [8]:

$$s_t^2 = \sum_i (\partial y/\partial x_i)^2 \sigma_i^2, \text{ or } s_t = \sqrt{\left(\frac{Obs}{Act}\right)^2 \sigma_{Obs}^2 + \left(\frac{res}{Act}\right)^2 \sigma_{res}^2 + \left(\frac{BW}{Act}\right)^2 \sigma_{BW}^2 + \left(\frac{jitter}{Act}\right)^2 \sigma_{jitter}^2} \quad \text{Eq. 3}$$

where the partial derivatives are taken with respect to each term $x_i$ and the $\sigma_i$ are the corresponding errors for each term. One can reduce the contribution by either reducing the term value or its error. In the case of the bandwidth term, we used the long pass filter to reduce the term compared to the short pass filter. Additionally, we performed a number of tests to evaluate the effect with different filters and two streak ranges, and we simulated the chromatic dispersion to give us a BW error of 10%. This is the major contribution to our errors in general as seen in Fig. 4. Here there is a 60% relative error to a 1 ps FWHM pulse or 0.4 ps sigma pulse. The tube resolution only contributes a 10% relative error at these same points. If enough samples (25) are taken the statistical error is less than the BW error. The BW relative errors are clearly smaller at only 10 % for a bunch length of 2.35 ps (FWHM) or 1.0 ps (sigma).

## 2.2 STREAK CAMERA MIRROR INPUT OPTICS

Our next phase of tests followed the installation of the A6856 mirror input optics on the streak camera in the tunnel and the replacement of a final UV lens with a focusing mirror. These steps reduced the quartz path by 35-38 mm for the optics barrel, the major contributor to the dispersive effects, and 10 mm for the final lens. We did obtain a reduction of the BW effect, but it was not as large as we had estimated from simple optics considerations for the remaining quartz vacuum window and 5.02 m of air.

Simplistically, there is a reduction of the bandwidth effect:

1) We can acquire data with the mirror optics without using a bandpass filter. Formerly, we assigned 6 ps FWHM to this BW term with the UV lenses. It is now about 2.5 ps (subtracting in quadrature a BW term) as inferred by matching to the CTR interferometer-measured short-bunch results.

2) When we inserted the 550-nm long pass filter (LPF) in the optical path, this BW term with UV lenses was ~2.5 ps FWHM. With the mirror optics this term reduces to about 1.8 ps FWHM, a reduction of 25%. We had expected <0.5 ps as a final BW term based on optical path length reduction.

Both of these results are determined by comparing the camera data and the CTR interferometer data on a beam configuration that gives about a 1.1 ps FWHM pulse as we scan through a series of quadrupole settings during x-z emittance exchange operations. We suggest that there is still a fundamental chromatic effect within the streak tube that leads to a slight blur of the temporal profile. Basically, photoelectrons are generated at the photocathode with different kinetic energies, energy spreads, and mean velocities depending on the wavelength of the incident photons. As a consequence, one obtains better resolution for 800-nm photons than with 266-nm photons due to kinetic energy spread, and there are small transit-time effects as well within the tube we suspect. These would also be a factor with unfiltered broadband OTR.

## 2.3 Delta Function Response of a Streak Tube

More recently, a Ti:Sa laser system was added to our laser lab at the A0PI facility [10]. This allowed for generation of ultrashort pulses at 800 nm and control of the longitudinal properties with a DAZZLER system produced by Fastlite [11]. The approximately 50-100 fs (FWHM) pulses can be used to probe more accurately the delta-function-like response of a streak-camera system to narrowband light since the streak-tube photocathode has sufficient sensitivity at 800 nm for imaging. The calibration factor was checked on the laser lab streak camera by using both an optical etalon loaned to us by Hamamatsu and the Colby Model PDL-100A-10.0ns delay box as shown in Figs. 5 and 6 [12]. The two methods show a very good agreement with a deduced calibration factor of 0.155 and 0.154 ps/pixel, respectively, for Range 1. The ORCA readout camera had twice as many pixels as the Pulnix camera used with the same model streak camera in the tunnel described in Section 2.1.

We then used the DAZZLER to adjust the bunch length of the laser pulse and tracked this with both a frequency-resolved optical gate (FROG) method and the laser lab streak camera. We show in Fig. 7 that if we use the delta function response in this case subtracted in quadrature from the observed bunch length in the streak camera, we obtain reasonable agreement with the FROG results. It is noted that one has sensitivity, albeit reduced, to bunch lengths shorter than the resolution limit. We note the focus image size was smaller than the delta-function-response size which was used here. Care was needed in balancing the input signal for the 1Hz and 1 kHz repetition rates into the streak tube using neutral density (ND) filters due to potential, intrinsic space-charge effects, but we see reasonable agreement. The shifts in the streak camera curve minima and the FROG data are attributed to the addition of those ND filters in the setups which introduced additional chirp effectively. We also confirmed separately that the streak camera system in the tunnel has about 0.6 ps (sigma) delta-function response to the monochromatic IR laser pulse. We conclude ideally one would use a 100-fs-long OTR pulse in Section 2.1, but we could not generate one in this facility.

## 3. STREAK CAMERA OTR RESULTS

The experiments were usually initiated by verifying the OTR-deduced spot sizes and centering the beam on the screen centerline and the downstream rf BPM coordinates. We would optimize the signal transported through the entrance slit of the streak camera while in Focus mode. We then switched to either range 2 or range 1, set the delay for viewing the streak images, and phase locked the delay box.

**3.1 Early Experiments at X9**

In the case of the straight-ahead beamline, we first tried to use larger charges in the micropulse and integrated over 10 pulses. Figure 8 shows the results for both range 2 and range 1 with 5.3 nC per micropulse and 10 micropulses synchronously summed in the image. For these data we actually used a 550 x 40 nm bandpass filter so the bandwidth effects are negligible. After subtracting a larger limiting resolution of 9 pixels (FWHM) from each image for this setup, the bunch length was determined as 31± 2ps in range 2 and 32 ± 2 ps in range 1. This was a somewhat larger value than expected so we next did a series of bunch length measurements in which we varied the micropulse charge only. The drive laser bunch length was maintained at about 7.9 ps (FWHM), as verified by a separate streak camera measurement. The results are shown in Fig.

9 where a significant elongation of the e-beam micropulse from 10 to 30 ps occurred as we varied the charge/micropulse from 1 nC to 5 nC. For comparison we also show a previous measurement reported on this photoinjector which has a decidedly different slope [6]. We attribute this difference to the laser spot size being reported as 1.77 mm rms compared to the present measured value of 1.15 mm rms. Since the laser bunch lengths were within 10 % in each case, the space charge effects should be stronger in our data. This is consistent with the ASTRA results (black curve) that simulate our present situation and are shown in Fig. 9 as well. The previous ASTRA simulation curve falls very close to those reported experimental data [6].

As part of these studies we also noted an apparent beam size variation *within* the micropulse time scale as shown in Fig. 10 for a micropulse charge of 4 nC. The observed synchronous sum beam size in the center region of interest (ROI) is about 34.5 pixels (FWHM), while it is 50.4 pixels in the upper ROI. A beam-size ratio of 1.46 for that portion in the upper ROI over the center ROI was found. The bunch length is about 28 ps (FWHM).

**3.2 Experiments in the EEX beamline**

The next series of investigations was done in the other beam line that is setup for emittance exchange experiments [2,3]. A liquid-$N_2$-cooled 5-cell $TM_{110}$ rf deflector cavity is positioned between two transport doglegs. Compression of the e-beam can be done by proper phasing of the upstream 9-cell cavity. In Fig. 11 we show the results of our initial experiments. The 9-cell rf accelerator phase is varied from 5 degrees off crest to about 25 degrees off crest [13]. The on-crest phase is 148 degrees in this case. The minimum bunch length is expected at about 21 degrees off crest based on simulations. Due to lower OTR signal transport, we used 100 micropulses of 1 nC per micropulse summed in the images. The bunch length is seen to vary from 22 ps down to 3.8 ps (FWHM) for these conditions. We used both R2 and R1 with the 550 nm SP filter, and then one additional scan another day was done with the 550 nm LP filter in place with R1. The data were matched in phase at the local minimum. The overlap of points for the different ranges and filters are all in good agreement. This supports the validity of our analysis of the various contributions to resolution in this regime.

Also the trajectory change with beam energy can be studied via the transit time changes through the doglegs as shown in Fig. 12. The phase-locked streak images allow the change in image time position to be used to track the arrival time change. One can see that a ±1% change in momentum causes about a ±6 ps change, respectively, in transit time through the bends. These data were used to evaluate one of the transport matrix elements of the emittance exchange line [2].

As a demonstration of the effects of the EEX beamline on the bunch length, we compared the observed streak images with the $TM_{110}$ cavity power off and at 100 % on as shown in Fig. 13. Due to the exchange of the incoming transverse emittance to the outgoing longitudinal emittance, the bunch length is clearly reduced from 2.4 ps (σ) to about 0.7 ps (σ). Each datum is a synchronous sum of micropulses over one macropulse running at 1Hz. So the plots show the scatter within the 100 shot samples in each state with the 5-cell cavity power alternating between on and off about every 100 seconds or macropulses.

An additional check of our streak camera image analysis was provided by a direct comparison to the bunch lengths determined from a Martin-Puplett interferometer which also viewed the X24 beamline location. In this case FIR CTR was used to assess bunch length [14], and we compare rms bunch lengths from each system as shown in Fig. 14 as we varied the upstream quadrupole before the EEX line. This quadrupole-field change affects the transverse beam size entering the EEX line, and it thus results in a bunch length variation in the outgoing beam. For rms bunch lengths shorter than 1 ps, we would expect the interferometer to be a reference. However, one sees that the interferometer data do not track the streak camera results to longer bunch lengths. We attribute this underestimation effect to the low-frequency cutoff of 0.10 THz in the interferometer system. The effect has been qualitatively reproduced using an autocorrelation code [15], and the calculated underestimate approaches 25% at 2.35 ps for 0.10 THz and 100% for 0.25 THz case compared to the actual bunch length as shown in Fig. 15.

As one of the demonstrations of the use of the mirror input optics, we probed the limits of resolution by using compression by the EEX beamline. In this case the variation of the current in the upstream quadrupole Q2AX06 (denoted as Q3 in Fig. 1) produces changes in the transverse x size which are exchanged to the bunch length after the EEX line. The data as shown in Fig. 16 are obtained at X24 using a micropulse charge of 250 pC and compared to simulations. A sum over 50 micropulses was used so there are some contributions from the intramacropulse jitter of the drive laser and the streak camera sweep. We assign 200 fs to each of these contributions so the lowest bunch length observed is close to the quadrature sum value, although the simulation predicts shorter bunches could be generated. The bandwidth term for broadband OTR is also a contributor to this experimental result.

### 3.3 Initial Laser and e-beam Synchronization Tests

More recently, as part of the preparations for a series of electro-optic sampling (EOS) experiments [9], we arranged for the transports of the drive laser UV component at 263 nm, the Ti:Sapph laser IR component at 800 nm, and the OTR from X9 to the entrance slit of the in-tunnel streak camera. An example of the "simultaneous" imaging of the three sources is shown in Fig. 17. Due to the synchroscan operations, the images could appear without all being on the same sweep, but they did appear in the similar relative phase positions to the sinusoidal sweep. We developed our techniques for changing the relative delay time between the sources so we could overlap them if desired. The Gaussian fits to the projected profiles are given in parentheses for the UV pulse ($\sigma$= 2.9 ps) which is generally shorter than the e-beam pulse ($\sigma$= 3.4 ps) due to space charge effects at 400 pC, and the IR pulse ($\sigma$= 0.9 ps) is basically a delta function test of the streak tube. This simultaneous imaging facilitated the sorting of laser and e-beam relative jitter effects and led to the implementation of a drive-laser-phase-feedback loop. The techniques were subsequently used for direct timing of the IR laser and the OTR at the X24 station for setting the relative timing to <1 ps for the EOS experiments [16].

## 4. DUAL-SWEEP STREAK-CAMERA RESULTS

With the addition of a dual-sweep (horizontal) deflection unit (Hamamatsu M5679), we have investigated the UV laser phase stability (both slew and jitter) during the macropulse and the electron beam's phase stability at the sub-ps level. Due to the intrinsic time delay in the horizontal deflection circuit, we display in Fig. 18 images of the last 19 micropulses of the 30 micropulses generated in each laser macropulse. On the vertical display axis is the fast R1 synchroscan time axis, and the horizontal display axis is the slow time sweep axis. The centroid of each micropulse image on the fast time axis (vertical) was determined as shown in Fig. 19a. A variance of 0.89 pixels (<300fs) in the presence of a small fitted phase slew is seen. An instrument image tilt (such as caused by the deflection plates not being exactly orthogonal to the readout CCD camera rows) is also evaluated by turning off the vertical deflection plates while running the horizontal sweep. The image tilt is shown as the dashed line in Fig. 19 b, and it accounts for 80% of the total time tilt observed. We estimate that the streak camera jitter within the macropulse contributes ~200 fs to the reported nominal jitter of 200 fs in the UV laser to give the total <300-fs result. These jitter values are about an order of magnitude smaller than those in reference [1], partly due to use of the newer C5680 camera and partly due to laser phase stability in this generation.

The final test involved looking at the intermacropulse jitter on the 20-s time scale. As shown in Fig. 20, the phase position is quite stable for the 4-micropulse average from point #2-11, but variations of about 1 ps then occur in the last 9 s of the sampling range. A more complete tracking study of the phase stability with a feedback loop added on the laser is given in reference [14].

## 5. SUMMARY

In summary, we have extended the investigations on streak camera imaging of the 15-MeV electron beam in the transport lines of the A0 photoinjector using OTR as the conversion mechanism. The enabling technology for these measurements was the synchroscan module installed in the streak camera mainframe combined with the new phase-locked delay box. These allowed synchronous summing of micropulses with much lower jitter than the single sweep unit and with the phase stability locked over 10s of minutes. We were then able to perform a comprehensive set of measurements on chromatic temporal dispersion effects in our optics, space-charge effects in the gun, bunch-compression effects, transit-time effects, the synchronization of laser and e-beam pulses, and the tracking of phase-jitter aspects. We also implemented the mirror input optics to reduce the bandwidth effects in OTR imaging and performed measurements with improved overall resolution as tested with the Ti:Sa laser pulses. A dual-sweep unit for the streak camera was added to the options to explore submacropulse time-scale effects at the 200-fs jitter level. The final upgraded system and techniques will be transferred to the new Advanced Superconducting Test Accelerator at Fermilab in the coming year.

## 6. ACKNOWLEDGEMENTS

The authors acknowledge support from M. Wendt, M. Church, and H. Edwards of Fermilab and A0 technical assistance from J. Santucci, R. Fliller, T. Koeth, A. Johnson, C. Tan, and M. Davidsaver. They acknowledge T. Maxwell and P. Piot for FROG Measurements and the Ti:Sa laser operation. They also acknowledge W. Cieslik of Hamamatsu Photonics for the loan of the optical etalon, for responses to many of our questions, and for first pointing out to one of us (AHL) the studies on input optics chromatic temporal dispersion effects many years ago.

## 8. FIGURE CAPTIONS

Figure 1: A schematic of the A0 photoinjector test area showing the PC rf gun, 9-cell Tesla booster cavity, transverse emittance stations, the OTR stations, the streak camera, and the EEX beamline when the two dogleg's dipoles are powered.

Figure 2: Calibration of the streak camera a) Range 2 and b) Range 1 using the laser pulse stacker. The separation of the split laser beam pulses was adjusted with the mirror spacing.

Figure 3: A simple representation of the group velocity dispersion effect on the streak image for a 3-ps FWHM initial pulse and a 4-ps temporal shift in the bandwidth used. The dotted curves represent the series of Gaussians time shifted with wavelength. The resultant streak image profile has a 4.21 ps (FWHM) size (solid curve).

Figure 4: Calculation of the relative error contributions to the FWHM for the dispersion/bandwidth effect and the tube resolution using Eq. 3. The bandwidth term is clearly the dominant factor for broadband sources.

Figure 5. The etalon calibration image (top) from the streak camera with ORCA CCD readout camera for Range 1 in synchroscan mode. The projected profiles (dots) on the time axis are shown in the lower plot with the Gaussian fits to the peaks (solid line). A calibration factor of 0.155 ps/pixel was obtained [12].

Figure 6. Plot of the laser lab streak camera image peak positions for 800 nm with the ORCA readout camera using the Colby delay box. The calibration factor in this case is 0.154 ps/pixel [12].

Figure 7. A comparison of the streak camera results and the FROG results (diamonds) as the 800-nm laser pulse rms bunch length is varied with different chirp settings for 1 Hz (crosses) and 1 kHz (circles) pulse repetition rates.

Figure 8: Streak camera image at 5.3 nC in the micropulses using a) range 2 and b) range 1. In the images the vertical axis is the time axis, and the horizontal display axis is the x spatial axis. The projected bunch length profile for each image is below it and has a width of 32±2 ps (FWHM).

Figure 9: Comparison of the variation of bunch length FWHM with micropulse charge as measured (squares) and as simulated by ASTRA (solid circles) and a previous measurement (diamonds) [6].

Figure 10: Evidence for slice beam-size effects on the micropulse time scale for a 4-nC micropulse charge. The x beam size sampled at each end of the time profile is about 50% larger than that sampled in the middle ROI of the pulse.

Figure 11: The variation of bunch length with 9-cell phase as measured by the streak camera after the two transport doglegs for two different ranges and the two filter options. At this level good agreement is seen.

Figure 12: A plot of the change in transit time through the doglegs for different 9-cell rf amplitudes, and hence momentum changes, Delta P.

Figure 13: Graphic demonstration of the shortened bunch lengths following emittance exchange with rf deflecting cavity ON compared to cavity OFF and with a charge of 1 nC [11].

Figure 14: Comparison of the streak camera and FIR interferometer results for quartz optics with systematic errors added to the statistical errors shown. The Q2AX06 (Q3 in Fig. 1) quadrupole current was scanned, and the bunch length varied via the EEX process. The charge was 250pC per bunch.

Figure 15: Calculated effects of two, low-frequency cutoffs of 0.10 (squares) and 0.25 THz (diamonds) on FIR autocorrelations and the determined bunch length (underestimated) versus the actual bunch length (solid circles).

Figure 16: Comparison of measured bunch lengths from the streak camera using mirror optics with simulations of EEX compression due to upstream Q2AX06 quadrupole (Q3 in Fig. 1) field variation.

Figure 17: Simultaneous recording of the synchroscan streak images for the OTR at X9, Ti:Sa IR, and UV drive laser (upper) and the time profiles with the Gaussian fits to the data (lower). Thirty bunches from the drive laser and OTR were used with a single IR pulse.

Figure 18: Example of dual-sweep streak images of the UV drive laser using R1 on the vertical deflection axis (10 ps is indicated for scale) and 20-µs coverage on the horizontal axis. The micropulses are separated by 1 µs in the macropulse.

Figure 19: Initial tracking of the drive laser's UV-component phase stability over 20 µs using R1-20 µs sweeps: a) linear slew fit to the data points and b) comparison to identified image tilt in instrument.

Figure 20: Initial tracking of the drive laser UV component's shot-to-shot macropulse phase jitter and slew. The dashed line is just for a reference in the phase-stable regime.

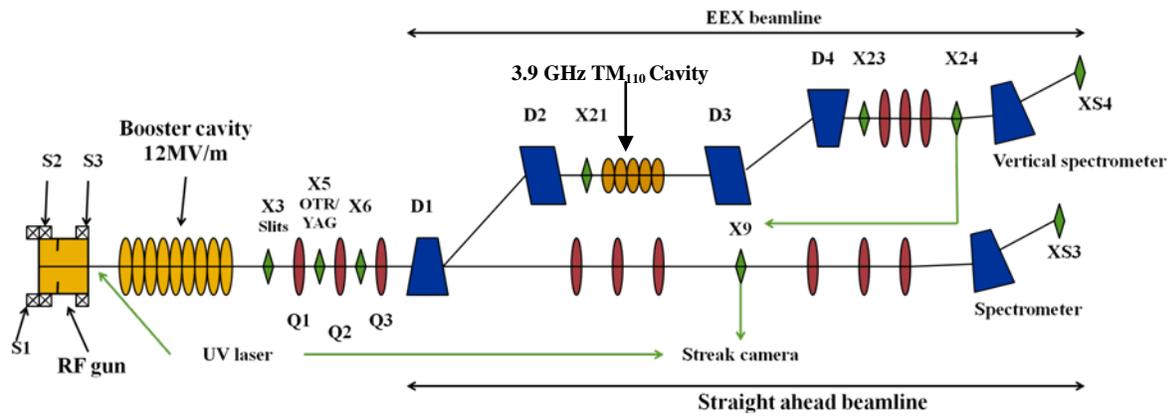

Fig. 1

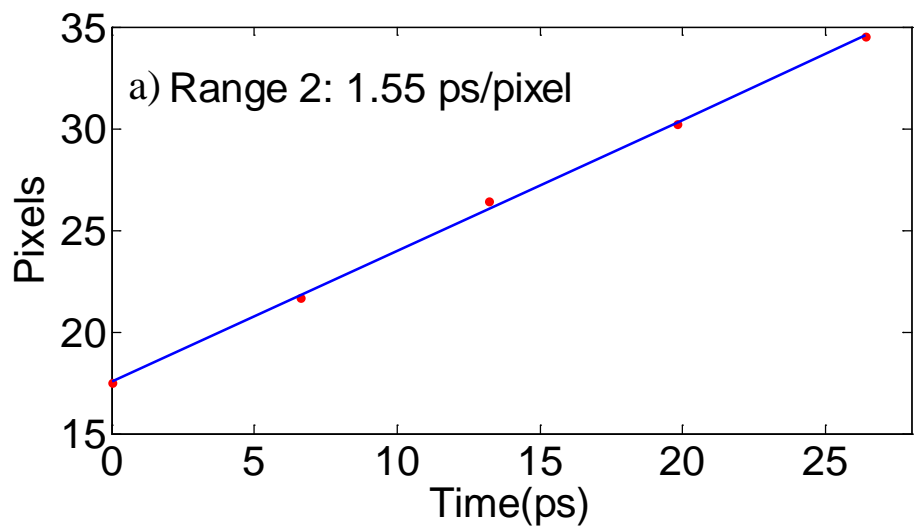

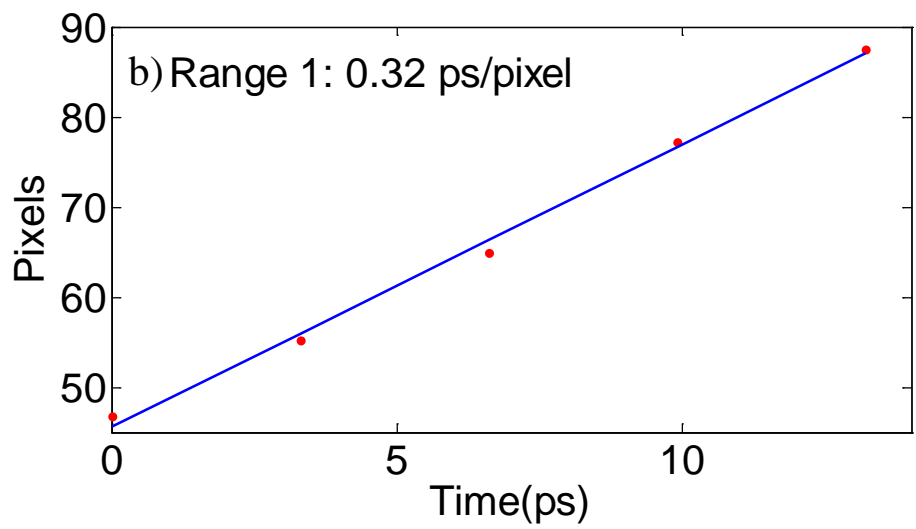

Fig. 2

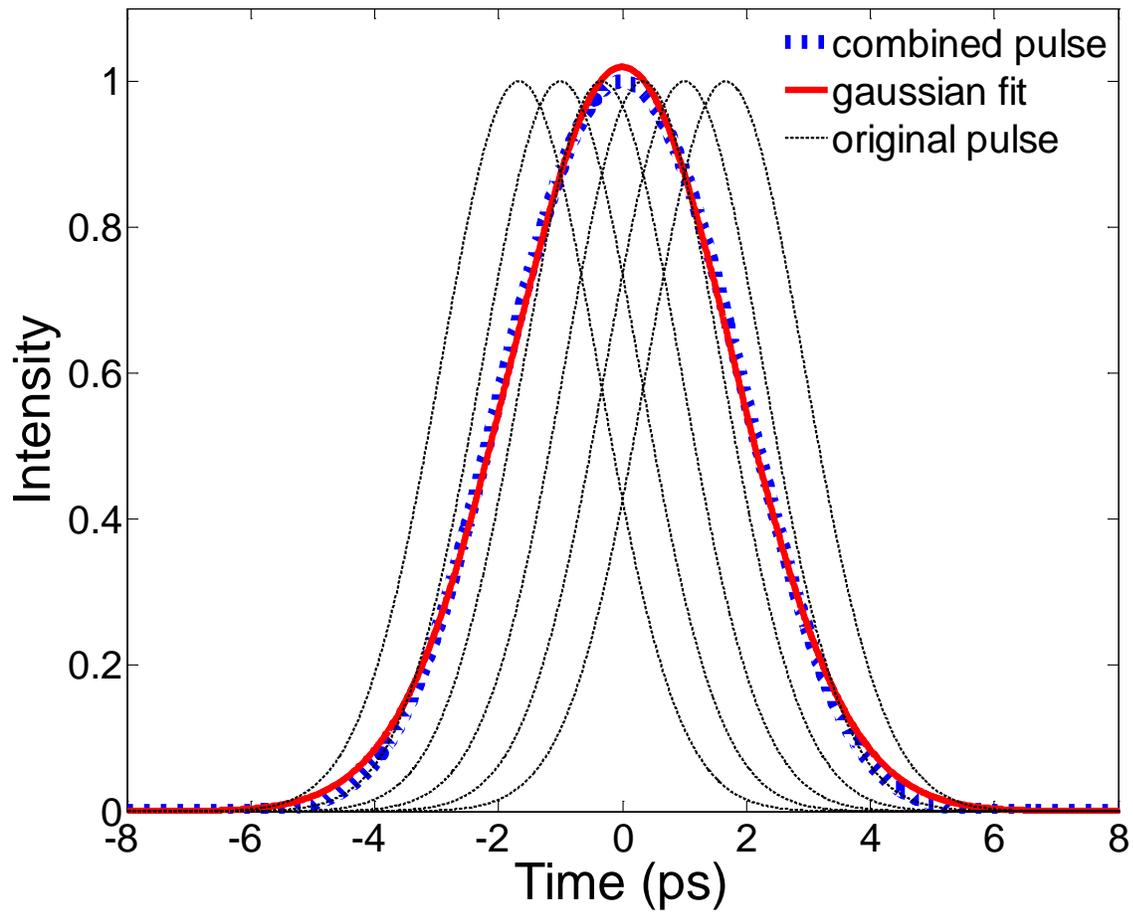

Fig. 3

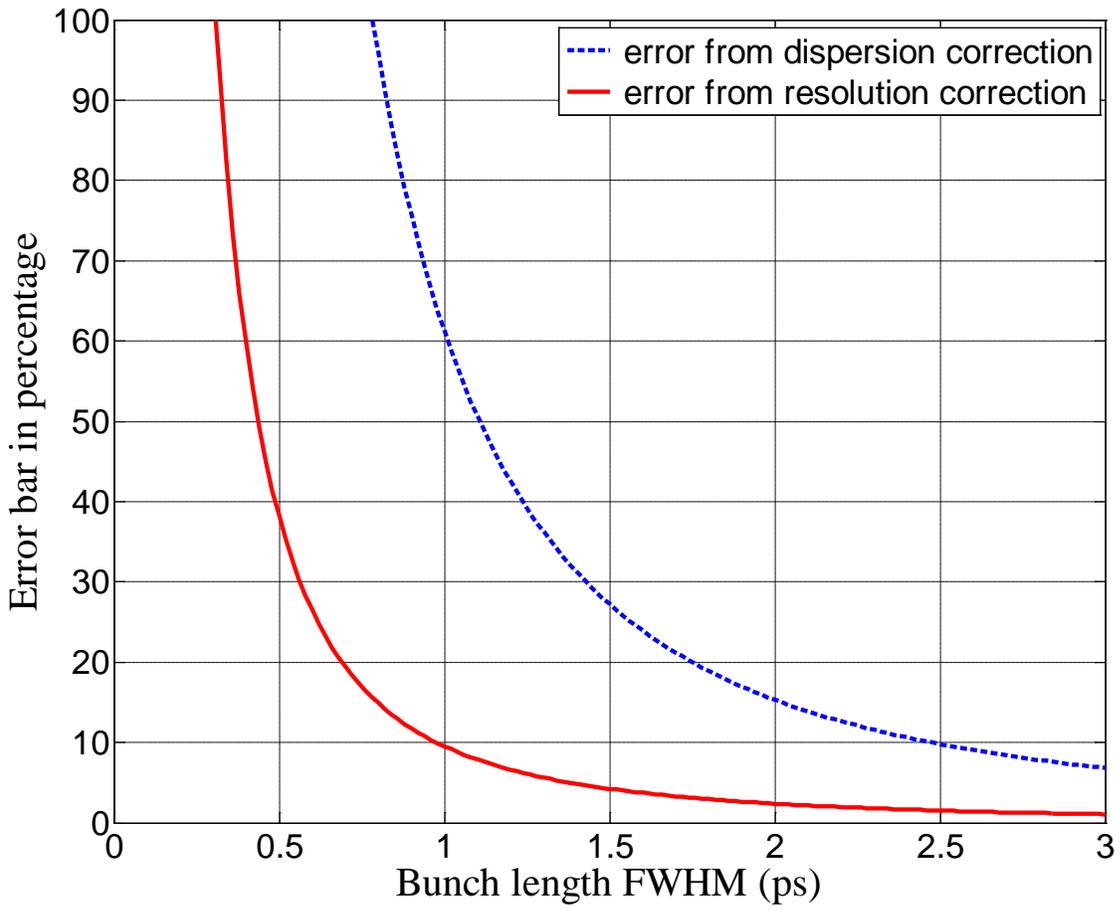

Fig. 4

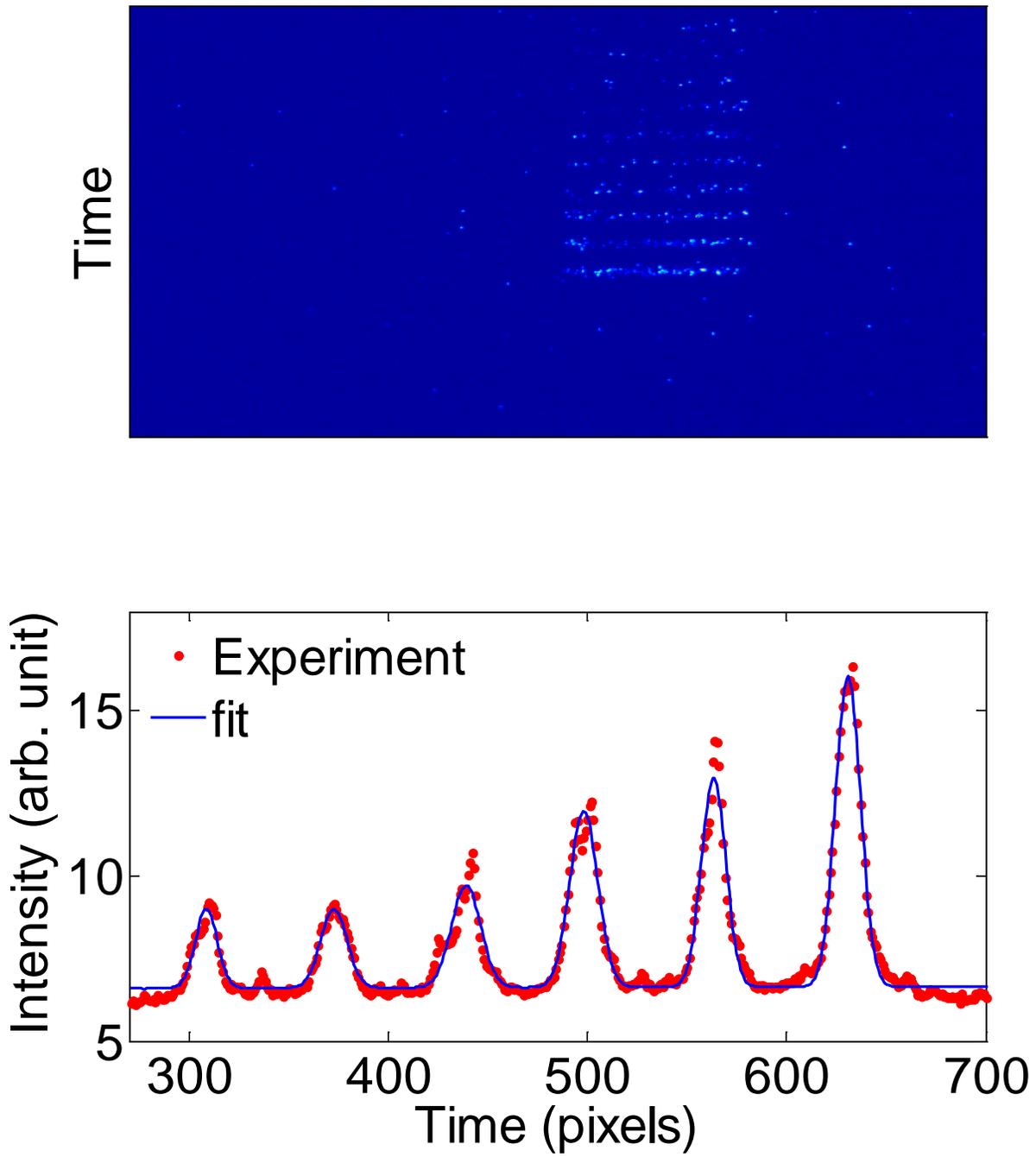

Fig. 5

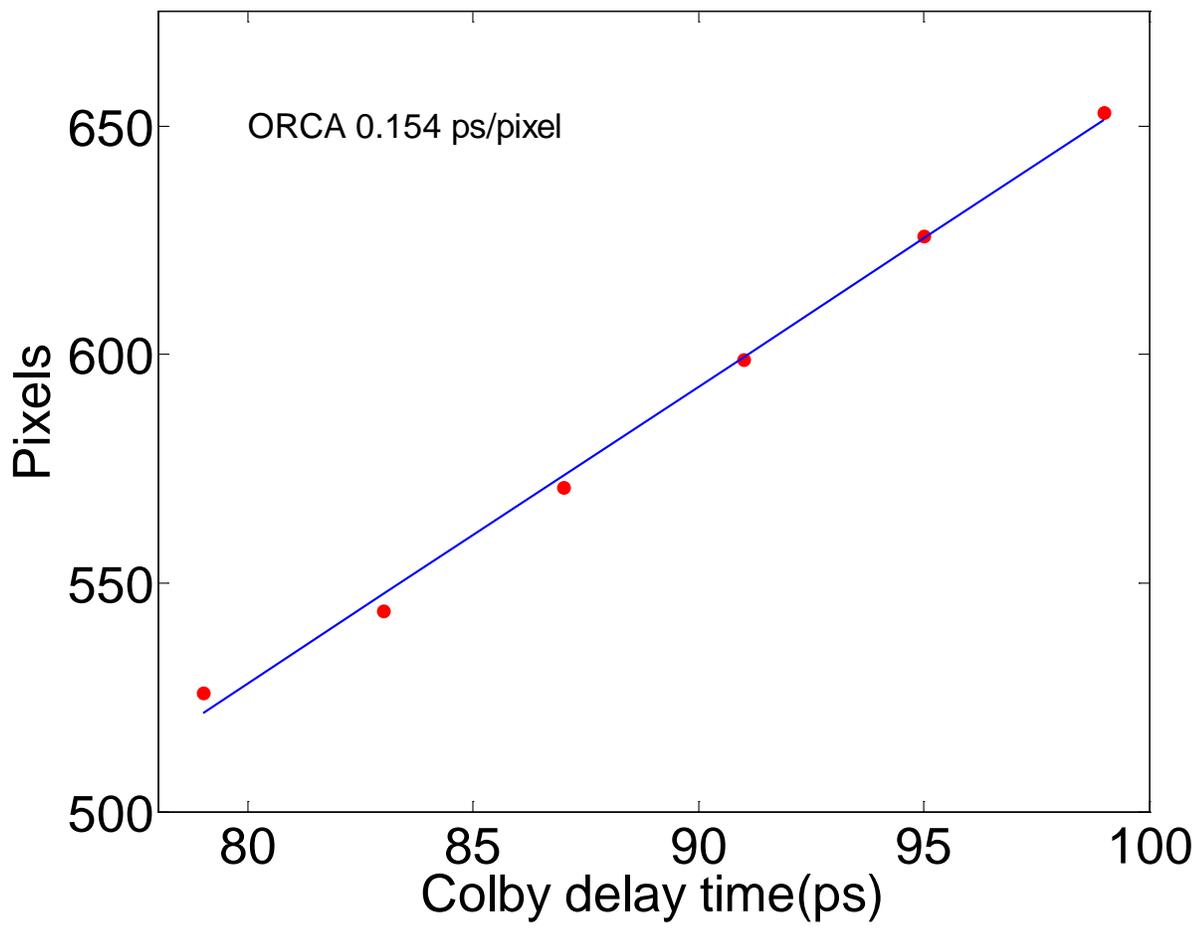

Fig. 6

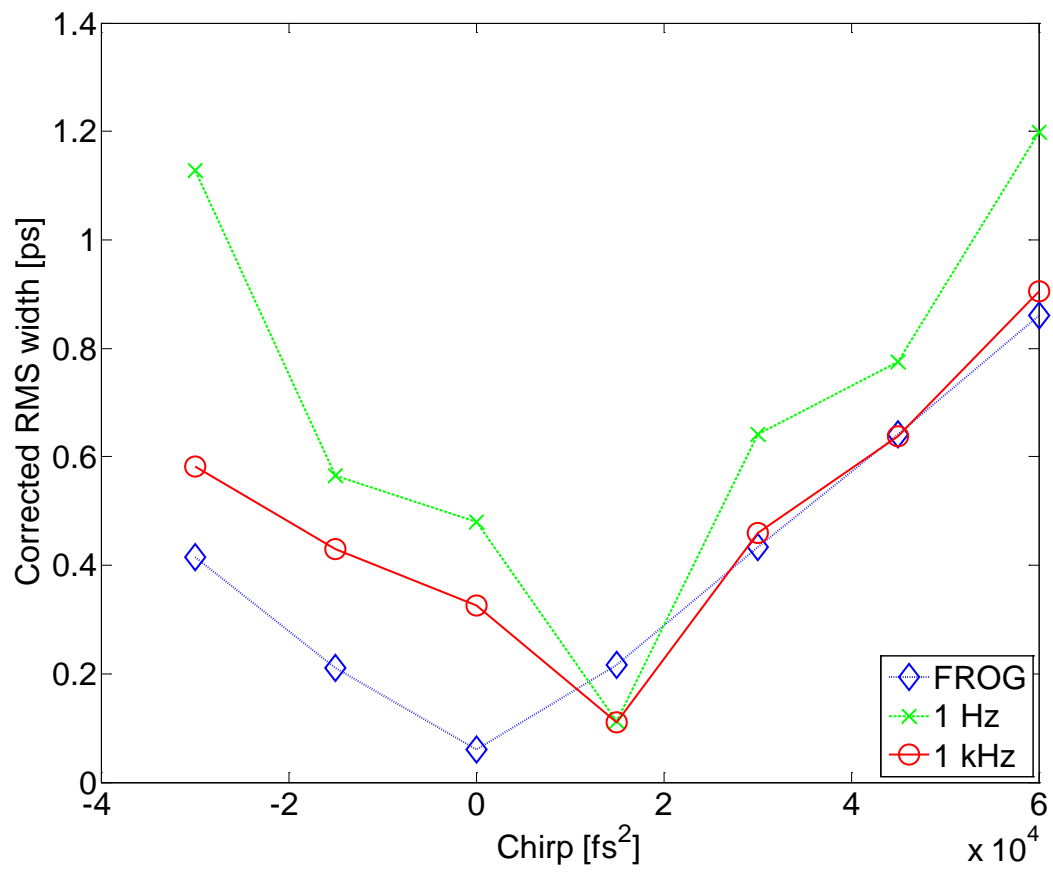

Fig. 7

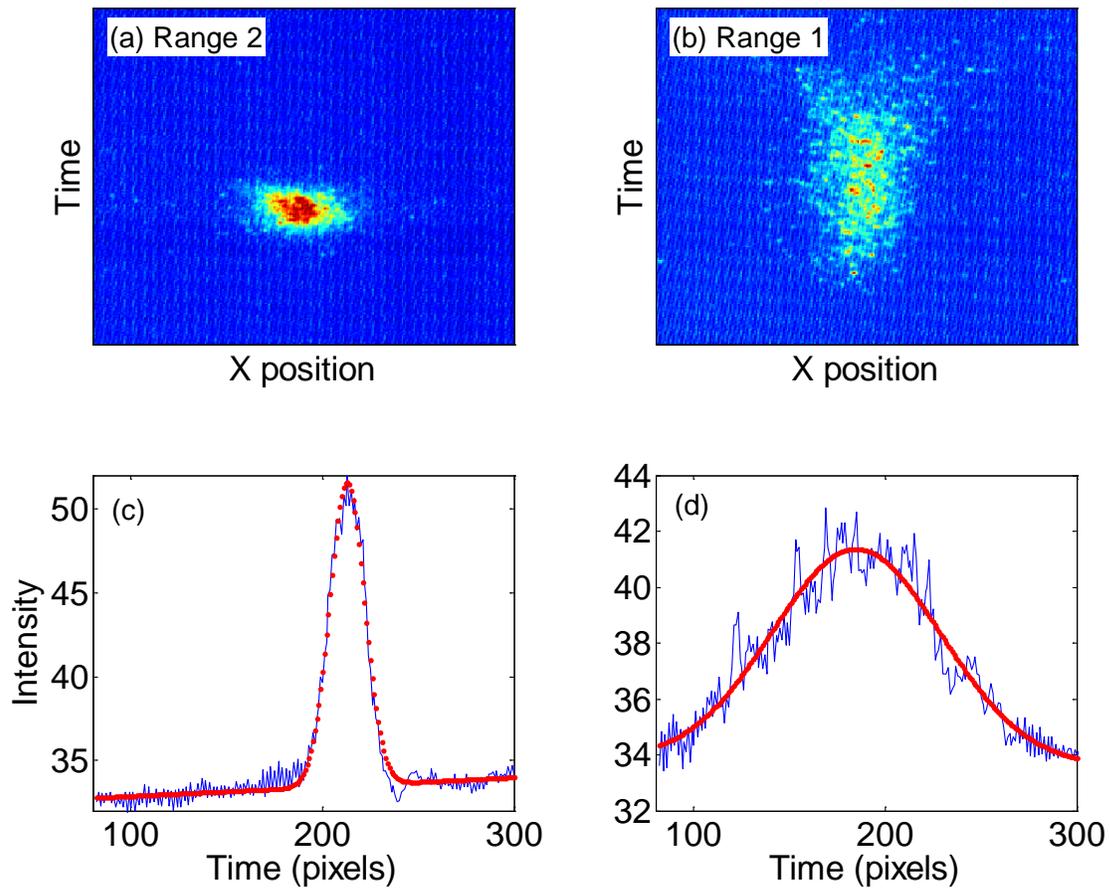

Fig. 8

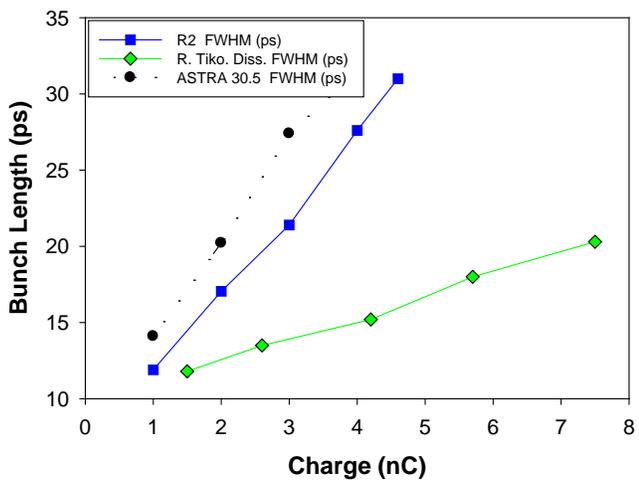

Fig. 9

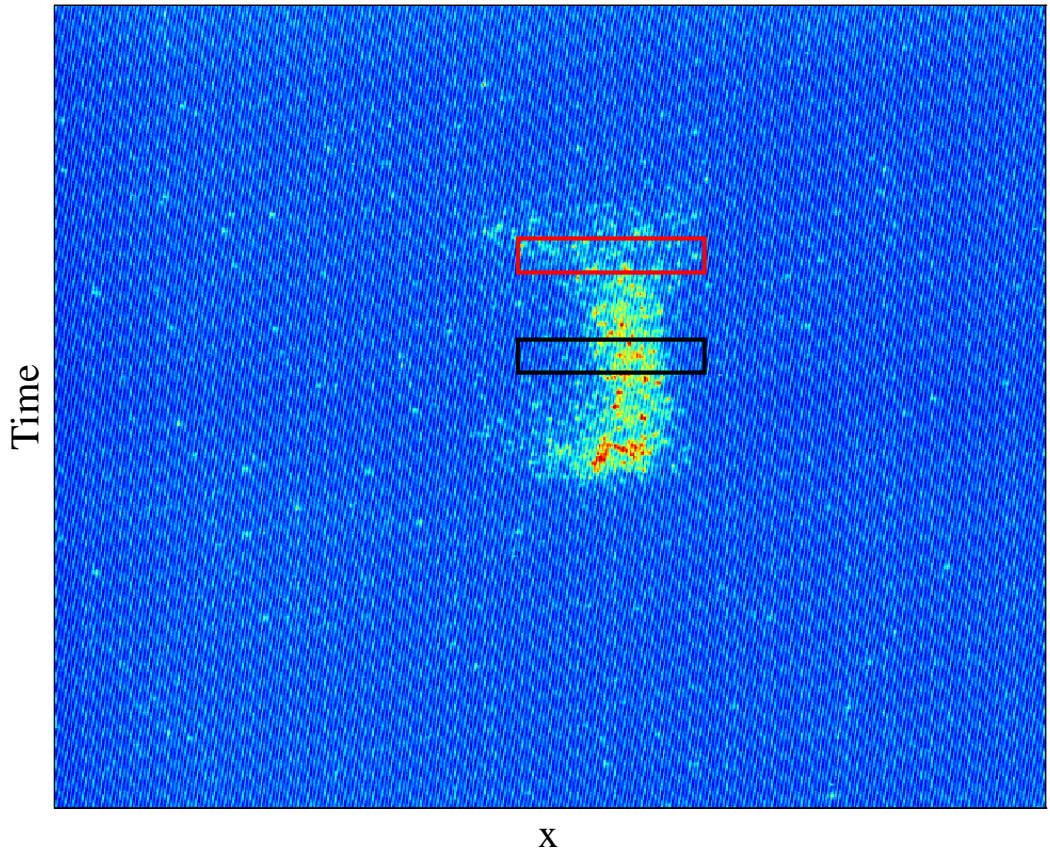

Fig. 10

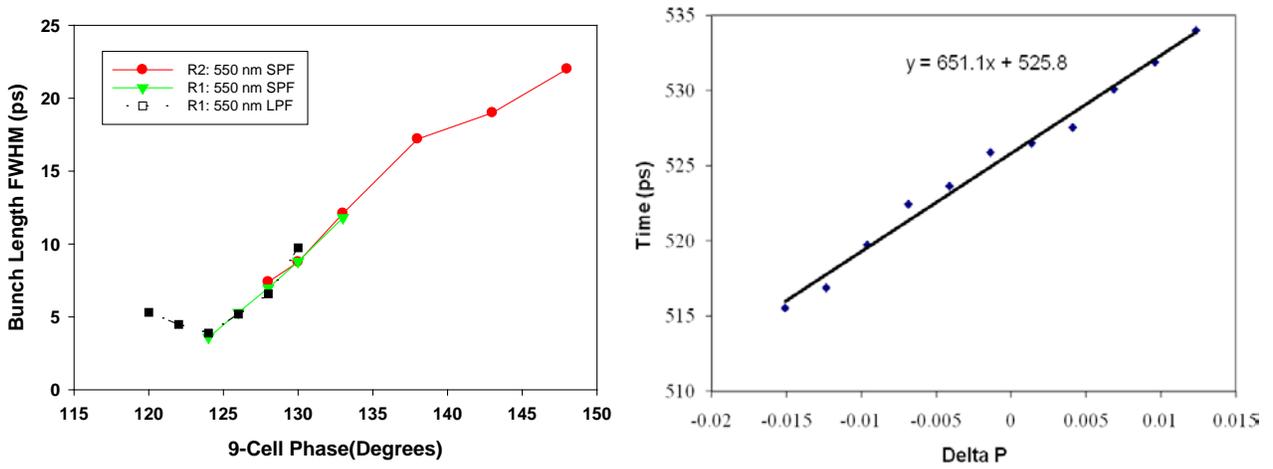

Fig. 11 Fig. 12

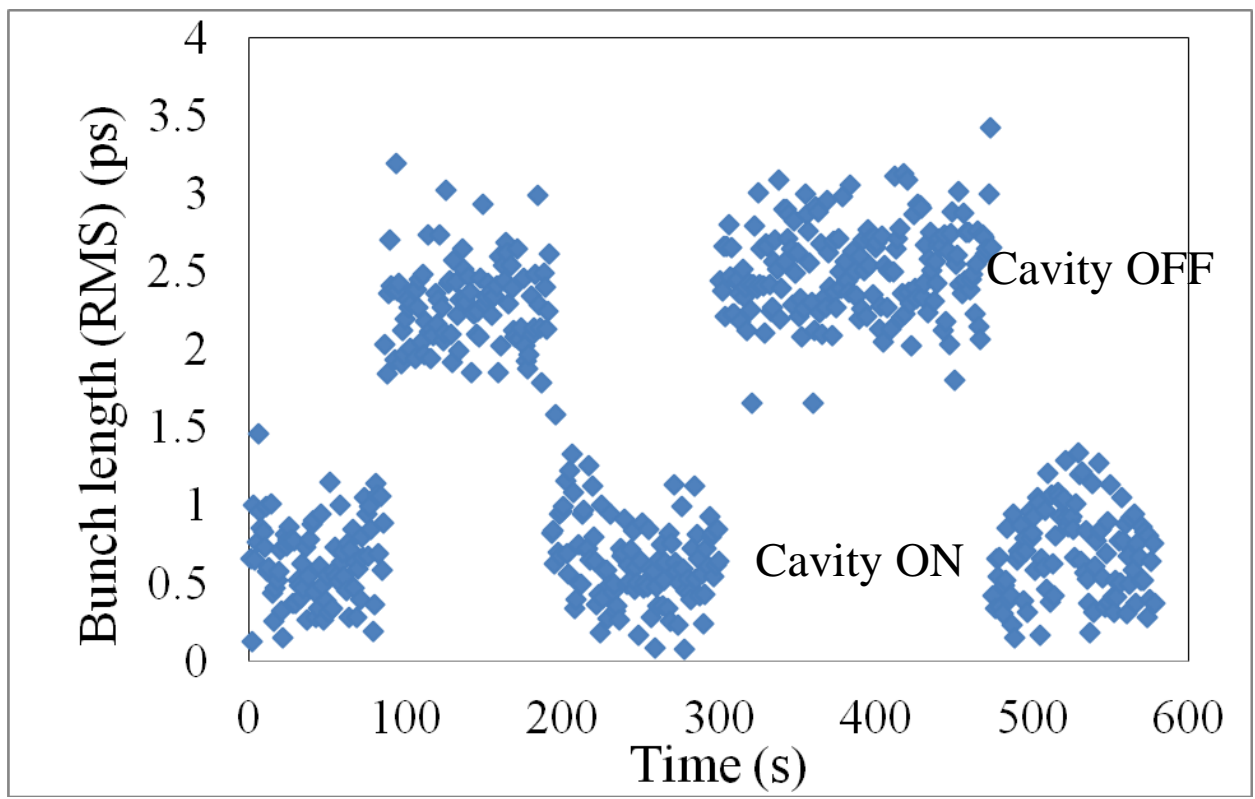

Fig. 13

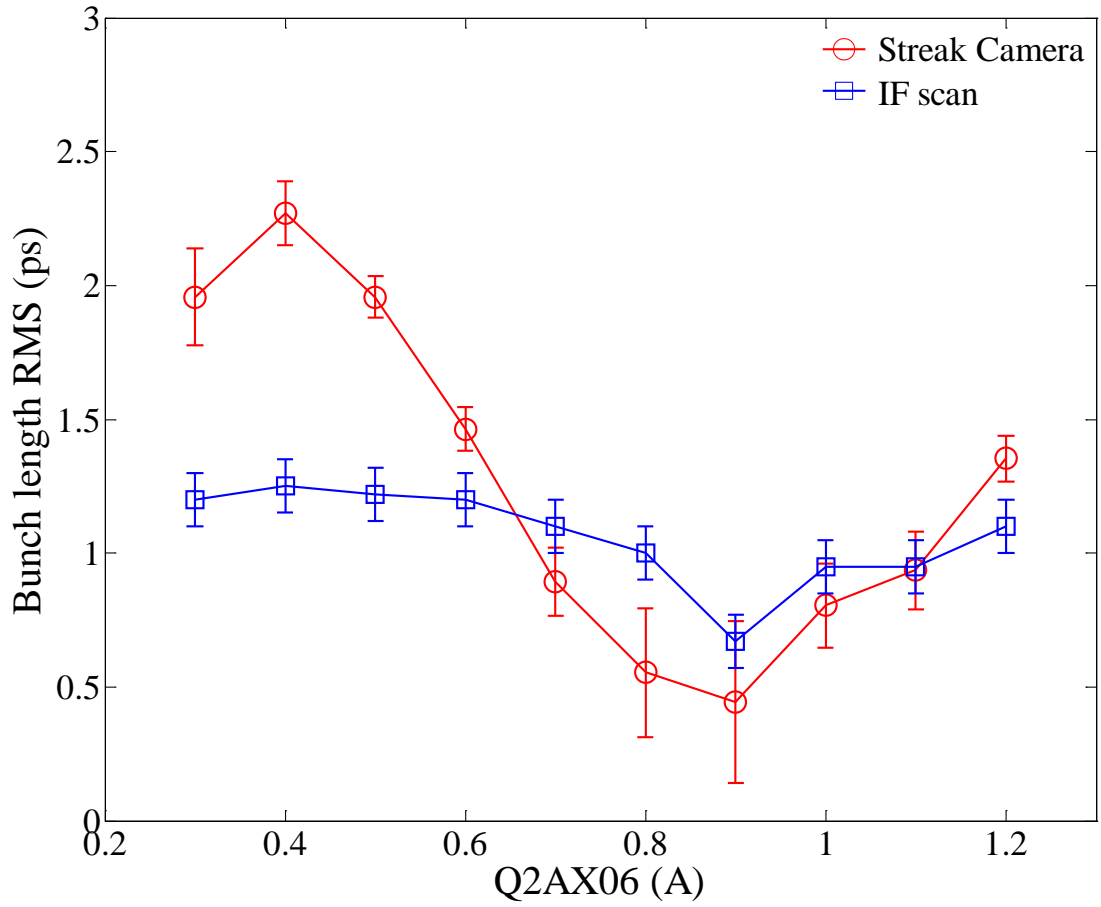

Fig. 14

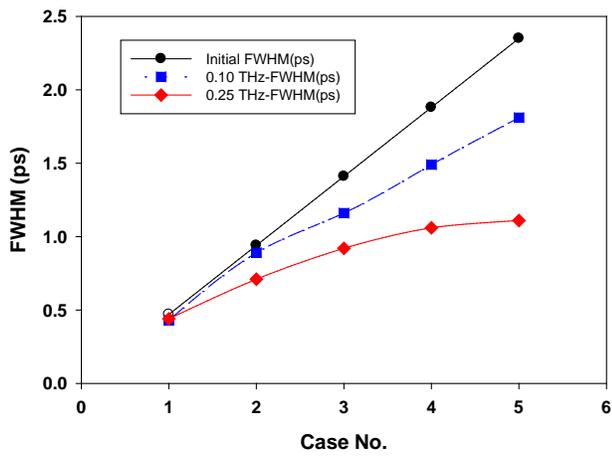

Fig. 15

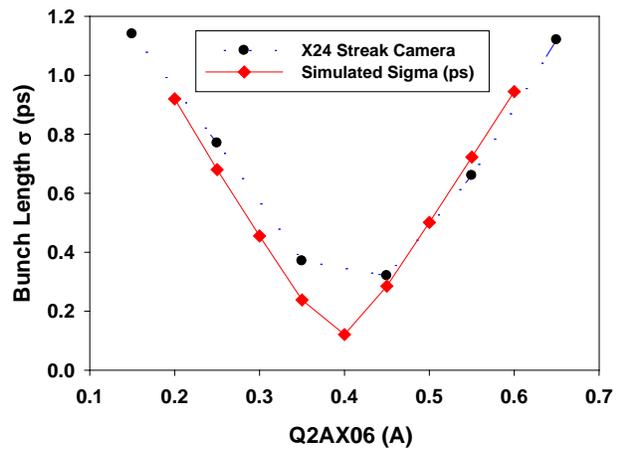

Fig. 16

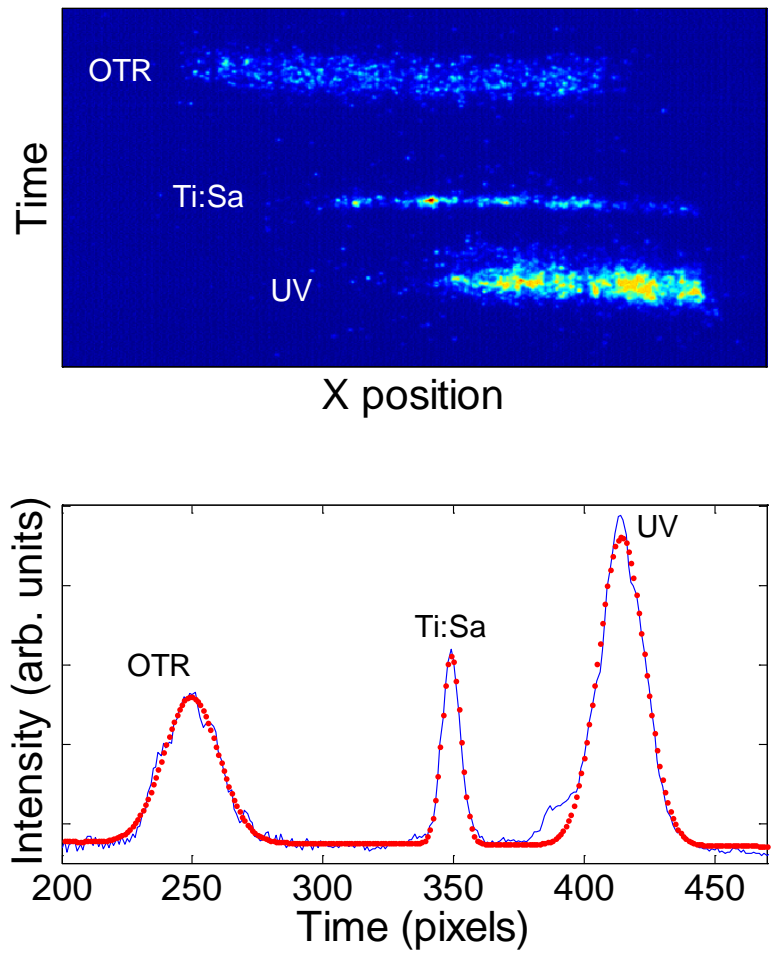

Fig. 17.

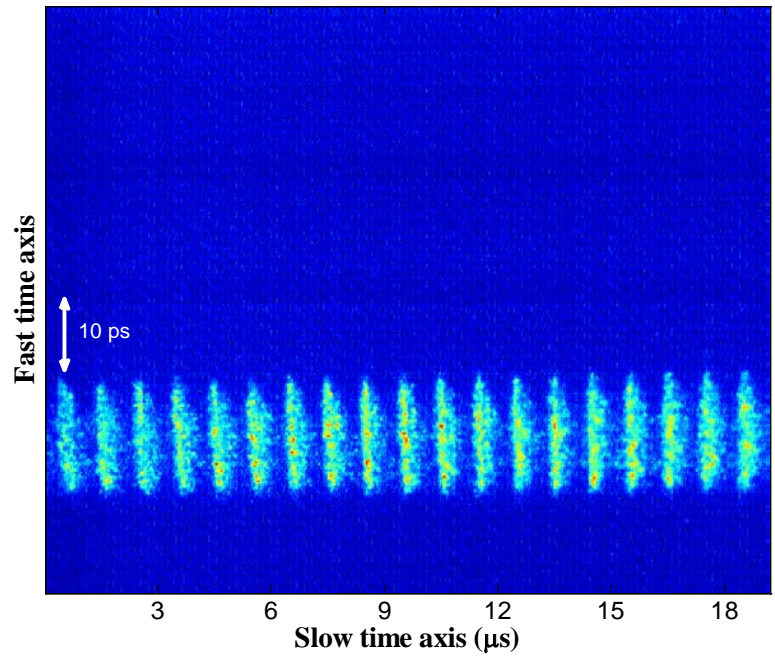

Fig. 18

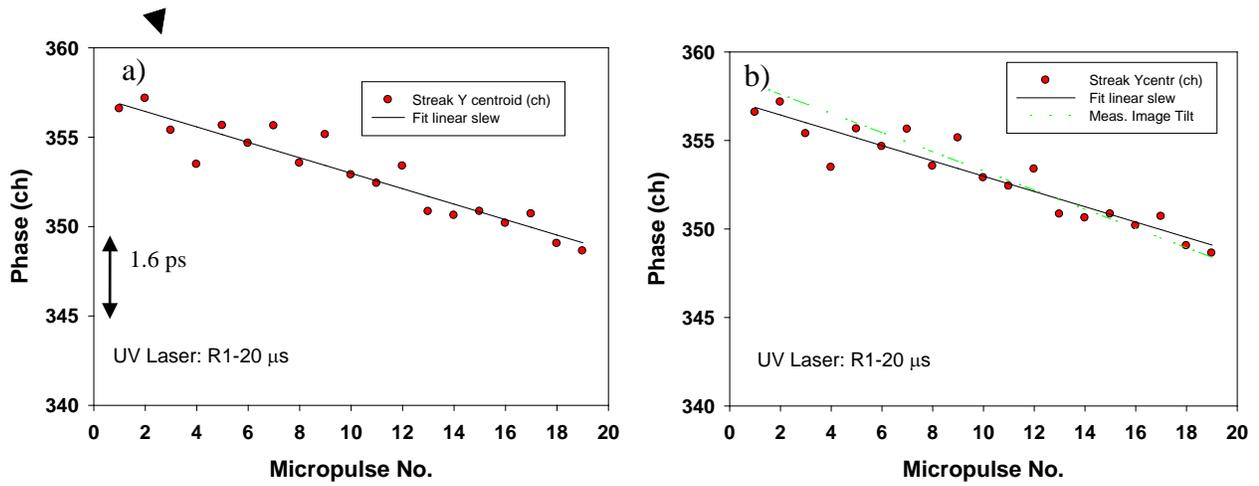

Fig. 19

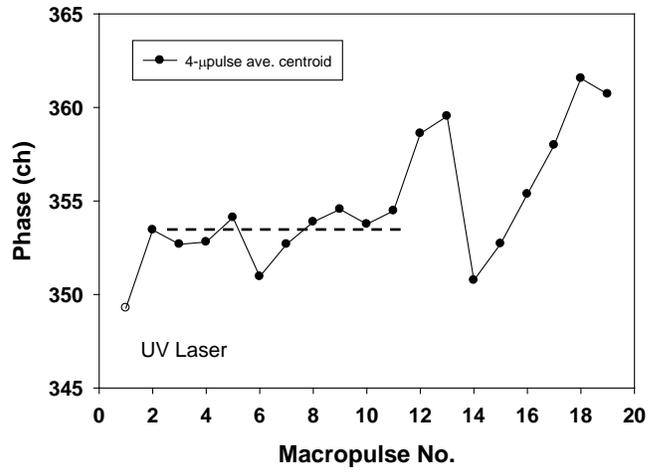

Fig. 20